\documentclass[% must pick either reprint or preprint
preprintnumbers,
nofootinbib,
superscriptaddress,
reprint,
amsmath,amssymb,
aps,
prc,
]{revtex4-2}

\usepackage{slashed} % For writing Feynman slash notation
\usepackage{microtype} % Fixes small typography errors by Latex
\usepackage{dcolumn}% Align table columns on decimal point
\usepackage{hyperref}% add hypertext capabilities

\usepackage{graphicx}% Include figure files

\begin{document}
\preprint{HUPD-2502}
%\title{Investigating electrical conductivity of quark-gluon plasma using a\\relativistic resistive magneto-hydrodynamic model}% Force line breaks with \\
% \title{Investigating the effects of QCD matter's electrical conductivity\\on charge dependent directed flow}
\title{Investigating effects of the electrical conductivity of QCD matter\\on charge-dependent directed flow}

\author{Nicholas J. Benoit}
\email{njbenoit@hiroshima-u.ac.jp}
\affiliation{Physics Program, Graduate School of Advanced Science and Engineering, Hiroshima University, Higashi-Hiroshima 739-8511, Japan}

\author{Takahiro Miyoshi}
\affiliation{Physics Program, Graduate School of Advanced Science and Engineering, Hiroshima University, Higashi-Hiroshima 739-8511, Japan}

\author{Chiho Nonaka}
\email{nchiho@hiroshima-u.ac.jp}
\affiliation{Physics Program, Graduate School of Advanced Science and Engineering, Hiroshima University, Higashi-Hiroshima 739-8511, Japan}
\affiliation{Department of Physics, Nagoya University, Nagoya 464-8602, Japan}
\affiliation{Kobayashi Maskawa Institute, Nagoya University, Nagoya 464-8602, Japan}
\affiliation{International Institute for Sustainability with Knotted Chiral Meta Matter, Hiroshima University, Higashi-Hiroshima 739-8511, Japan}

\author{Hiroyuki R. Takahashi}
\affiliation{Department of Natural Sciences, Faculty of Arts and Sciences, Komazawa University, Tokyo 154-8525, Japan}

\date{\today}% It is always \today, today,

\begin{abstract}
    Charge dependent directed flow is an important observable of electromagnetic fields in relativistic heavy-ion collisions.
    We demonstrate how the difference in charge dependent directed flows between protons and antiprotons is sensitive to the resistivity, inverse of quark-gluon plasma's electric conductivity, over different collision centralities.
    Our model numerically solves the 3+1D relativistic resistive magneto-hydrodynamic (RRMHD) equations, assuming the electric conductivity to be a scalar.
    For this work, we focus on symmetric Au + Au collisions at the top RHIC energy of $\sqrt{s}=200$~GeV.
    We illustrate the time evolution of the electromagnetic fields in our model and connect that to the charge dependent directed flow results.
    Our results highlight the importance of modeling quark-gluon plasma's electric conductivity for charge dependent observables in relativistic heavy-ion collisions.
\end{abstract}

\maketitle
%\tableofcontents

%%%% Introduction %%%%
\section{Introduction}

At high temperatures, hadronic matter smoothly transitions into a strongly interacting quark-gluon plasma (QGP).
Relativistic heavy-ion collision experiments at the Relativistic Heavy Ion Collider (RHIC) and the Large Hadron Collider (LHC) have produced and continue to investigate that deconfined hadronic state of QGP.
Experimental evidence favors a strongly coupled fluid behavior for QGP, which leads to phenomenological models of the QGP expansion using relativistic viscous hydrodynamics~\cite{jacak_exploration_2012}.

Simultaneously, the relativistic charges of the collision spectators create intense electromagnetic (EM) fields in and around the collision region.
Estimates using Li\'enard-Wiechert potentials place the initial magnetic field for RHIC energies of $\sqrt{s}=200$~GeV at $e|\vec{B}| \sim \mathcal{O}(m^2_\pi)$ and LHC energies of $\sqrt{s}=2760$~GeV at $e|\vec{B}| \sim \mathcal{O}(10m^2_\pi)$~\cite{bzdak_event-by-event_2012,deng_event-by-event_2012}.
Because of the collision geometry, those intense EM fields penetrate the created QGP, which modifies the life-time of the EM fields.
Lattice QCD, perturbative QCD (pQCD), and AdS/CFT calculations predict QGP to be a resistive medium~\cite{hattori_electrical_2016,li_conductivities_2018,astrakhantsev_lattice_2020,aarts_electrical_2021,ghosh_electrical_2024,almirante_electrical_2024}.
Using that information, the interplay between currents in QGP and the EM fields can be parametrized by the electric conductivity $\sigma$ transport coefficient.

A few collision observables have also been suggested to investigate the electric conductivity of QGP, including using electroweak probes~\cite{yin_electrical_2014,rapp_electric_2024}.
An example relevant to this paper is the charge dependent directed flow in asymmetric collisions~\cite{hirono_estimation_2014}.
Theoretically, in asymmetric collisions, an electric field is directed from the larger ion to the smaller ion because of the difference in the number of protons.
Then the electric field induces a charged current inside QGP via the Coulomb effect that modifies the directed flow of final state electrically charged hadrons.
The argument is that the difference between the directed flow of electrically charged hadrons $\Delta v_1 \equiv v_1^+ - v_1^-$ is sensitive to the electric conductivity of QGP.
At RHIC, the STAR detector has reported a difference in the directed flow between negatively and positively charged hadrons for Cu-Au collisions at $\sqrt{s_\text{NN}}=200$ GeV~\cite{star_charge-dependent_2017}.
However, the uncertainty of the result is too large and the understanding of other non-EM effects is not complete enough for any definite statements to be made about QGP's conductivity.

Instead, we will consider the charge dependent directed flow $\Delta v_1$ of symmetric collisions.
Charge dependent directed flow was suggested to be a signal of electromagnetic fields in symmetric collisions~\cite{gursoy_magnetohydrodynamics_2014,gursoy_charge-dependent_2018}.
Recent STAR data for the directed flow of protons and kaons supports that theory~\cite{star_observation_2024}, but results for pions and D mesons indicate non-EM effects are important~\cite{STAR_beam-energy_2014,alice_probing_2020}.
Directed flow $v_1$ in symmetric collisions has also been studied using relativistic magneto-hydrodynamics~\cite{inghirami_numerical_2016,inghirami_magnetic_2020} and a resistive model was used to study the $\Delta v_1$ for pions~\cite{nakamura_charge-dependent_2023}.
The result of the resistive model illustrated the $\Delta v_1$ dependence on the electric conductivity of QGP.

For this work, we have improved our relativistic resistive magneto-hydrodynamic (RRMHD) model \cite{nakamura_relativistic_2023} to study different final state hadrons across collision centralities\footnote{Previously only peripheral centralities and final state pions were possible.}.
We aim to show the qualitative role of QGP's electric conductivity in $\Delta v_1$.
This contrasts with other works that have focused on baryon charges and currents to explain the STAR data~\cite{star_observation_2024}.
Those works have suggested the positive slopes of $\Delta v_1$ can be solely explained by the transported quark effect.
But, as was pointed out in Ref.~\cite{gursoy_charge-dependent_2018}, the initial protons in the collision mean the plasma has a net positive charge.
We will demonstrate with our RRMHD model that the initial positive charge can also explain positive slopes for $\Delta v_1$.
This demonstration should be thought as a first step because the initial non-equilibrium dynamics of QGP have not been settled, hence an exact quantitative role of the initial positive charges is difficult to establish.

Physically, an imbalance in $\Delta v_1 \neq 0$ is thought to be a caused by the Faraday induction + Coulomb effects and the Hall effect acting on charged particles inside QGP.
If the combined Faraday induction + Coulomb effects are larger than the Hall effect, $\Delta v_1$ will feature a negative slope over the mid-rapidity region~\cite{gursoy_charge-dependent_2018}.

In the next section, Sec.~\ref{sec:formalism}, we introduce our relativistic resistive magneto-hydrodynamic (RRMHD) model.
Then in Sec.~\ref{sec:magneticfield} we will discuss the time evolution of the magnetic field in our model.
Next, how we fix the initial conditions of the EM fields as solutions of Maxwell's equations in a medium and the initial conditions of QGP are discussed in Sec.~\ref{sec:numericalsetup}.
Results for the time evolution of the magnetic field are shown in Sec.~\ref{sec:MagResults}, and we discuss how the results are related to the initial conditions.
In Sec.~\ref{sec:chargeDependentFlow} we present the results for charge dependent directed flow using our RRMHD model, including details on the connection with charge accumulation on the freezeout hypersurface.
Our conclusions are summarized in Sec.~\ref{sec:conclusions} and we discuss the necessary future improvements to move this type of study from qualitative to quantitative.

%%%% Methods %%%%
\section{Hydrodynamic formalism of quark-gluon plasma with conductivity}\label{sec:formalism}

For this work, we have built upon the relativistic resistive magneto-hydrodynamic (RRMHD) model ~\cite{nakamura_charge-dependent_2023,nakamura_directed_2023,nakamura_relativistic_2023} such that we can study central collisions.
That model simultaneously solves the relativistic ideal-hydrodynamic equations and Maxwell's equations with Ohm's law as a current source.
Reviewing the relevant details, we start with the conservation laws of ideal-hydrodynamics:
\begin{align}
    \label{eq:baryoncurrent}
    \nabla_\mu N^\mu & = 0,
    \\
    \nabla_\mu T^{\mu\nu} & = 0.
\end{align}
Here $\nabla_\mu$ is the covariant derivative.
Then those ideal-hydrodynamic equations are augmented with Maxwell's equations,
\begin{align}
    \nabla_\mu F^{\mu\nu} & = - J^\nu,
    \\
    \frac{1}{2}\nabla_\mu \epsilon^{\mu\nu\rho\sigma}F_{\rho\sigma} & = 0,
\end{align}
where $\epsilon^{\mu\nu\rho\sigma} = \tilde{\epsilon}^{\mu\nu\rho\sigma}/\sqrt{-g} $ is the Levi-Civita tensor and tilde epsilon is the antisymmetric symbol.
Then to close the system of equations we use the first order covariant form of Ohm's law,
\begin{align}\label{eq:ohmslaw}
    J^\mu & = q u^\mu + \sigma F^{\mu\nu}u_\nu,
    % \\
    % & = q u^\mu + \sigma e^\mu
    \\ \label{eq:restframeEfield}
    & =  \gamma \begin{pmatrix} q + \sigma v_jE^j, & q v^i+ \sigma [E^i+\epsilon^{ijk}v_jB_k] \end{pmatrix},
\end{align}
and use the ideal equation of state (EoS).
We have assumed the electric conductivity $\sigma$ is a constant and $q=-J^\mu u_\mu$ is the electric charge density in the fluid's comoving frame.
Following the naming convention of Ref.~\cite{gursoy_charge-dependent_2018}, the $E^i$ of Eq.\eqref{eq:restframeEfield} is the Coulomb + Faraday currents and $\epsilon^{ijk}v_jB_k$ is the Lorentz current.

The defining feature of RRMHD compared to ideal relativistic magneto-hydrodynamics (MHD) is a finite conductivity $\sigma$, which is the inverse of the resistivity.
In ideal relativistic MHD models of QGP, the conductivity is taken to be infinite,  in other words, QGP is treated as a perfect conductor.
However, Lattice QCD and pQCD calculations suggest the conductivity should be $\sigma = \mathcal{O} (0.1 \sim 10)$ for the highest energy regions produced at RHIC and the LHC~\cite{aarts_electrical_2021}.
Additionally, a finite conductivity reduces the lifetime of the EM fields compared to ideal MHD and can introduce new phenomena like magnetic reconnection.
Modeling QGP with a finite conductivity leads to interesting physics beyond the ideal MHD scenario.

%%%% Discussion on Methods %%%%
\section{Time evolution of the magnetic field}\label{sec:magneticfield}

The magnetic field plays a major role in the charge-dependent directed flow $\Delta v_1$ at symmetric collisions.
One role the magnetic field plays is to induce Faraday and Lorentz currents in the QGP.
During the time evolution, the interplay of Lorentz + Coulomb + Faraday currents within the plasma leads to local imbalances in the electric chemical potential.
Then that finite chemical potential becomes frozen onto the freezeout hypersurface.
The local imbalances on the freezeout hypersurface caused by the time evolution of the magnetic field changes electric charge dependent observables like $\Delta v_1=v_1^+-v_1^-$.
Before discussing the changes to $\Delta v_1$, we will focus on the time evolution of the magnetic field in our RRMHD model.

Numerous calculations have been done to estimate the magnitude and lifetime of the magnetic field in relativistic heavy-ion collisions under simplifying assumptions~\cite{gursoy_charge-dependent_2018,gursoy_magnetohydrodynamics_2014,inghirami_magnetic_2020,inghirami_numerical_2016,tuchin_initial_2016,tuchin_time_2013}. 
In our model, we have three assumptions:
\begin{itemize}
    \item We assume the electric conductivity of QGP as a constant value over all spacetime.
    Strictly speaking, dimensional analysis alone suggests the electric conductivity is temperature-dependent, which means it should change as the plasma expands and cools.
    \item We ignore the pre-equilibrium dynamics that can create an initial electric charge distribution and change the configuration of the initial EM fields~\cite{yan_dynamical_2023,mclerran_comments_2014}.
    Instead, we have chosen to initialize the EM fields using the solutions for the in-medium Maxwell equations~\cite{tuchin_time_2013}.
    Those solutions have been shown to overestimate the EM field values~\cite{tuchin_initial_2016,yan_dynamical_2023,mclerran_comments_2014}.
    We have roughly estimated how that overestimation affects our model in Appendix \ref{append:initialEMfields}.
    \item Last, the initial EM fields are created by the spectators of the collision, but during the time evolution we only include the plasma charges as a source of the EM fields.
    In other words, our model focuses on the behavior of charges inside the plasma and how they alone affect the time evolution of the EM fields.
    This may lead to an underestimation of the late time behavior of the EM fields, which we discuss near the end of Sec.~\ref{sec:MagResults}.
    % However, we do not expect the small differences 
    % late time behavior the of the EM fields to have significant changes to experimental observables like $v_1$.
\end{itemize}
For a more complete description, we plan to resolve the first and third items in a future version of our model.
The second item depends on an understanding of the pre-equilibrium dynamics, which has recently been an active area of research (e.g., ~\cite{matsuda_simulation_2024,Li_2023}).
Some possible directions include color glass condensate (CGC), IP-Glasma, flux tube, and kinetic models.

With these assumptions in mind, our model captures how a finite electric conductivity modifies the time evolution of the magnetic field.
For an intuition of the modifications we derive the magnetic wave equation in Milne coordinates, starting from Faraday's and Ampere's laws:
\begin{align}
    \frac{1}{\tau}\tilde{\epsilon}^{ijk}\partial_j E_k + \frac{1}{\tau}\partial_0(\tau B^i) & =0,
    \label{eq:faraday}
    \\
    \frac{1}{\tau}\tilde{\epsilon}^{ijk}\partial_j B_k - \frac{1}{\tau}\partial(\tau E^i) & = j^i.
    \label{eq:ampere}
\end{align}
In the usual way, we take the $\partial_0$ of Faraday's Law and substitute it into Ampere's law to find a wave equation for the magnetic field components,
\begin{gather}
    \Box_M (\tau B^1) = 2\tau \partial_2 E^3 + \partial_3 j^2 - \tau^2 \partial_2 j^3,
    \\
    \Box_M (\tau B^2) = - 2\tau \partial_1 E^3 - \partial_3 j^1 + \tau^2 \partial_1 j^3,
    \\
    \Box_M (\tau B^2) = - \partial_1 j^2 + \tau^2 \partial_2 j^2, 
\end{gather}
where $\Box_M$ is the d'Alembert operator in Milne coordinates defined to be,
\begin{equation}
    \Box_M \equiv -\partial_0\partial_0 - \frac{1}{\tau}\partial_0 + \partial_1\partial_1 + \partial_2\partial_2 + \frac{1}{\tau^2}\partial_3\partial_3.
\end{equation}
Those are the same equations that have been discussed in the literature~\cite{dash_charged_2024}.
As introduced in Sec.~\ref{sec:formalism}, we augment Faraday's and Ampere's laws Eq.\eqref{eq:faraday} and Eq.\eqref{eq:ampere} with Eq.\eqref{eq:ohmslaw} Ohm's law.
Because our goal is a qualitative feel for the magnetic field's time evolution in Milne coordinates, we simplify Ohm's law by neglecting any particle-density gradients and considering the fluid's rest frame.
Then the diffusion current becomes $j^i = \sigma E^i$ and the magnetic field wave equations are,
\begin{gather}
    \Box_M (\tau B^1) = 2\tau \partial_2 E^3 - \sigma (\tau^2 \partial_2 E^3 - \partial_3 E^2),
    \\
    \Box_M (\tau B^2) = - 2\tau \partial_1 E^3 - \sigma (\partial_3 E^1 - \tau^2 \partial_1 E^3),
    \\
    \Box_M (\tau B^2) = - \sigma (\partial_1 E^2 - \partial_2 E^1).
\end{gather}
Since the sigma dependent terms on the right-hand side of those equations are Faraday's law, we rewrite them to arrive at:
\begin{align}
    \partial^\mu\partial_\mu(\tau B^1) - \left( \frac{1}{\tau} + \sigma \right) \partial_0(\tau B^1) & = -2\tau\partial_2 E^3,
    \label{eq:nonlinearBx}
    \\
    \partial^\mu\partial_\mu(\tau B^2) - \left( \frac{1}{\tau} + \sigma \right) \partial_0(\tau B^2) & = -2\tau\partial_1 E^3,
    \label{eq:nonlinearBy}
    \\
    \partial^\mu\partial_\mu(\tau B^3) - \left( \frac{1}{\tau} + \sigma \right) \partial_0(\tau B^3) & = 0.
    \label{eq:nonlinearBz}
\end{align}
Equation \eqref{eq:nonlinearBz} is for a non-linearly damped magnetic wave in the $\eta$ direction.
Equations \eqref{eq:nonlinearBx} and \eqref{eq:nonlinearBy} are similar to Eq.\eqref{eq:nonlinearBz} except for a driving term proportional to the electric field.
Even though those three equations are non-linearly damped, the damping is simple enough to gain a qualitative feeling of the magnetic field time evolution.
\begin{itemize}
    \item For $\sigma\geq 1$ the magnetic field is always over damped because the proper time is always greater than zero, $\tau>0$, in Milne coordinates.
    \item For $0<\sigma<1$ the nature of the magnetic field damping depends on time.
    Starting in the region $\tau < 1/(1-\sigma)$ the magnetic field is under-damped, then a smooth transition occurs to an over-damped region $\tau>1/(1-\sigma)$.
    The exact time of the transition depends on the value of $\sigma$.
\end{itemize}

%\begin{itemize}
%    \item For, $\tau^{-1}+\sigma > 1$ the magnetic field is over damped and exponentially decays.
%    \item At, $\tau^{-1}+\sigma = 1$ the magnetic field transitions between over and under damped.
%    \item Then for, $\tau^{-1}+\sigma < 1$ the magnetic field is under damped and will oscillate about an asymptotic value.
%\end{itemize}
% the magnetic field is always over damped when $\sigma > 1\text{ fm}^{-1}$ and a transition is never realized.
% But it is possible for the magnetic field to transition from under damped to over damped if the conductivity is between $0 < \sigma < 1$.
% Because of the non-linearity of the damping, it is possible for the magnetic field to transition from under damped to over damped if the conductivity is between $0 < \sigma < 1$.
% As an example, for, $\sigma = 0.0294\text{ fm}^{-1}$ the transition occurs close to $\tau = 1\text{ fm/c}$.
% For conductivities larger than 1 the transition would occur at negative $\tau$, which cannot be realized.
% Thus, the magnetic field is always over damped when $\sigma > 1\text{ fm}^{-1}$.
This will be illustrated in Fig.~\ref{fig:comparedTimeEvo} of Sec.~\ref{sec:MagResults}.

%%%% Methods %%%%
\section{Numerical setup and methods}\label{sec:numericalsetup}

For relativistic heavy-ion collisions, the symmetry of the system is best captured with Milne coordinates $(\tau, x, y, \eta)$, which have the metric $g_{\mu\nu}=\text{diag}( -1, 1, 1, \tau^2 )$.
Accordingly, our RRMHD model solves Eqs.\eqref{eq:baryoncurrent}-\eqref{eq:ohmslaw} using those coordinates with the volume factor $\sqrt{-g} = \tau$.
We have adopted the optical Glauber model for the initial energy density profile and included an initial longitudinal tilt~\cite{bozek_directed_2010}.
Even though we initialize the model after QGP is created at $\tau_0 = 0.4$ fm/$c$, the source of the initial EM fields is dominated by the collisions spectators~\cite{tuchin_initial_2016}.
Therefore, we calculate the initial EM fields using the analytic solution from~\cite{tuchin_time_2013}, which is a solution of Maxwell's equations in a medium with the ion charges as the source.
The calculation is summarized in Appendix~\ref{append:initialconditions}.
Overall, these are the same initial conditions as the previous works~\cite{nakamura_directed_2023,nakamura_charge-dependent_2023,inghirami_magnetic_2020} and are listed in table~\ref{tab:initialparameters}.
\begin{table}[hbt]
\centering
    \begin{tabular}{|c|c|}
         & RHIC \\
        \hline
        Ion & $^{197}$Au \\
        $\sqrt{s_\text{NN}}$ & $200 \text{ GeV}$ \\
        $\tau_0$ & $0.4$ fm/$c$ \\
        $\epsilon_0$ & $55.0 \text{ GeV/fm}^3$\\
        $\sigma_\text{NN}$ & $41.3$ mb \\
        $\eta_s$ & $5.9$ fm  \\
        $\omega_\eta$ & $0.4$  \\
        $\epsilon_\text{vac}$ & $0.10 \text{ GeV/fm}^3$
    \end{tabular}
    \caption{Initial parameters used for this work from Refs.~\cite{nakamura_directed_2023,nakamura_charge-dependent_2023,inghirami_magnetic_2020}.}
    \label{tab:initialparameters}
\end{table}
The inelastic cross-section $\sigma_\text{NN}$ is calculated with the parametrization $\sigma_\text{NN}(s) = 28.84 + 0.0458 \ln^{2.374}[s]$ from Ref.~\cite{denterria_progress_2021}.

Numerically, we use a constant grid size of $168 \times 168 \times 72$ with $-16.8\text{ fm} < (x,y) < 16.8\text{ fm}$ and $-7.2 < \eta < 7.2$, except for calculating the initial EM fields.
Instead, for the initial EM fields, we adopt the following approach based on the analytic solution~\cite{tuchin_time_2013}, which is summarized in Appendix~\ref{append:initialconditions}.
\begin{enumerate}
    \item Assume the nuclei are very thin in the beam direction $z$ because of Lorentz contraction, such that their charge is distributed only in the transverse plane $x, y$.
    Then we can fix the $z$ position of the charges based on the collision kinematics.
    \item At a grid point $\begin{pmatrix} x, & y, & \eta \end{pmatrix}$, calculate the EM fields by integrating over the contribution from all possible sources.
    Practically, this means we create a new grid for the EM sources in the transverse plane of size $200 \times 200$ for $-15.0\text{ fm} < (x, y) < 15.0 \text{ fm}$ and numerically integrate over it.
    This size ensures the area of both nuclei are integrated over.
\end{enumerate} 
The second step is repeated for all the grid points $\begin{pmatrix} x, & y, & \eta \end{pmatrix}$.
For the numerical time evolution, we have adopted $C_\text{CFL}=0.05$ for the Courant-Friedrichs-Lewy (CFL) value.

Once the time evolution is completed, we calculate the final state hadron distribution using the Cooper-Frye formula~\cite{Cooper_1974}.
\begin{equation}\label{eq:cooperfrye}
  \frac{d N}{d Y p_T d p_T d \phi} = \int_{\Sigma_f} f (x, p) p^{\mu} d\Sigma_{\mu}.
\end{equation}
We are ignoring any final state interactions that would be included in an afterburner because our current goal is to focus on the effects of the QGP conductivity.
Our implementation of the Cooper-Frye formula constructs a 4-D hypersurface at the energy density boundary $\epsilon_f = 0.15 \text{ GeV/fm}^3$~\cite{huovinen_particlization_2012}.
We construct the local distribution $f(x,p)$ of Eq.\eqref{eq:cooperfrye} from the fluid velocities $v_i$, and charge density $q$ that is perpendicular to the hypersurface $d\Sigma(t,x,y,z)$,
\begin{equation}\label{eq:localdistribution}
  f (x, p) = \frac{1}{2 \pi} \frac{1}{e^{[p_{\mu} u^{\mu} (x) - \mu_Q (x)
  Q] / T_f (x)} \mp 1},
\end{equation}
where $Q$ is the quantized electric charge of the hadron.

For this work, the time evolution of the fluid is solved using an ideal-gas equation of state (EoS)
\footnote{
We have assumed the chemical potential $\mu_Q (x)$ is much smaller than the temperature of the fluid.
Our assumption was based on the lack of baryon stopping at high-energy collisions observed near mid-rapidity and relating $\mu_Q = \mu_B/3$~\cite{star_identified_2004,star_systematic_2009}.
However, to mimic the experimentally observed splitting between the direct flow of hadrons, our model needs to include a finite chemical potential on the freezeout hypersurface.
Then, to construct the potential on the hypersurface we need to introduce a new EoS that depends on $\mu_Q$.
}.
However, on the freezeout hypersurface the distribution of Eq.~\eqref{eq:localdistribution} requires knowledge of the electric charge chemical potential $\mu_Q (x)$.
To calculate $\mu_Q (x)$, we relate it to the local charge density $q(x)$ using a different EoS, i.e., Ref.~\cite{monnai_four-dimensional_2024}, with the fluid's local pressure $P$ and the thermodynamic relation $q=(\partial P/\partial\mu_Q)$.
Up to the linear approximation of $q(x)$ the relation is,
\begin{equation}
  \mu_Q (x) \simeq \frac{q (x)}{g_h T_f^2}
\end{equation}
where $g_h$ is the hadronic degrees of freedom.
The final temperature in Eq.~\eqref{eq:localdistribution} is calculated for a massless boson gas,
\begin{equation}
    T_f = \left(\frac{30\epsilon_f(\hbar c)^ 3}{9\pi^2}\right)^{1/4} \approx 0.140 \text{ GeV,}
\end{equation}
using the boson gas degrees of freedom.
Finally, we integrate over the entire hypersurface to arrive at the final state hadron distribution.

%%%% Results/Discussions part 1 %%%%
\section{RRMHD Results: Time evolution of the magnetic field}\label{sec:MagResults}

As discussed at the beginning of Sec.~\ref{sec:magneticfield}, the time evolution of the magnetic field induces Lorentz + Faraday currents in QGP.
Coulomb currents caused by the electric field of the positive spectators also appear, but for symmetric collisions largely cancel each other.
Then the EM field component in the charge dependent directed flow $\Delta v_1$ for symmetric collisions is dominated by the Lorentz + Faraday currents Eq.\eqref{eq:restframeEfield}, specifically the magnetic field.
In particular, the $B_y$ component of the field has the largest contribution to $\Delta v_1$, which approximately appears as $\gamma B_y$ in the Faraday current and $-\gamma u_z B_y$ in the Lorentz current~\cite{gursoy_charge-dependent_2018}.
Where $\gamma$ is the Lorentz factor of QGP in the local fluid rest frame and $u_z$ is the $z$ component of the fluid velocity. 

In our RRMHD model, the time evolution of $B_y$ depends on the initial conditions and the electric conductivity $\sigma$ of QGP.
As a starting point, we have used smooth initial solutions to Maxwell's equations in a medium as discussed in Sec.~\ref{sec:numericalsetup} that predicts a large initial field strength for $B_y$.
% Because the initial dynamics of QGP are yet to be agreed upon, we save an analysis of their effects on $B_y$ for future work.
That being said, solving the time evolution of the magnetohydrodynamics Eqs.~\eqref{eq:baryoncurrent}-\eqref{eq:ohmslaw} with different $\sigma$'s has an impact on $\Delta v_1$ through changes of $B_y$.

\begin{figure}[bh]
    \includegraphics[width=\linewidth]{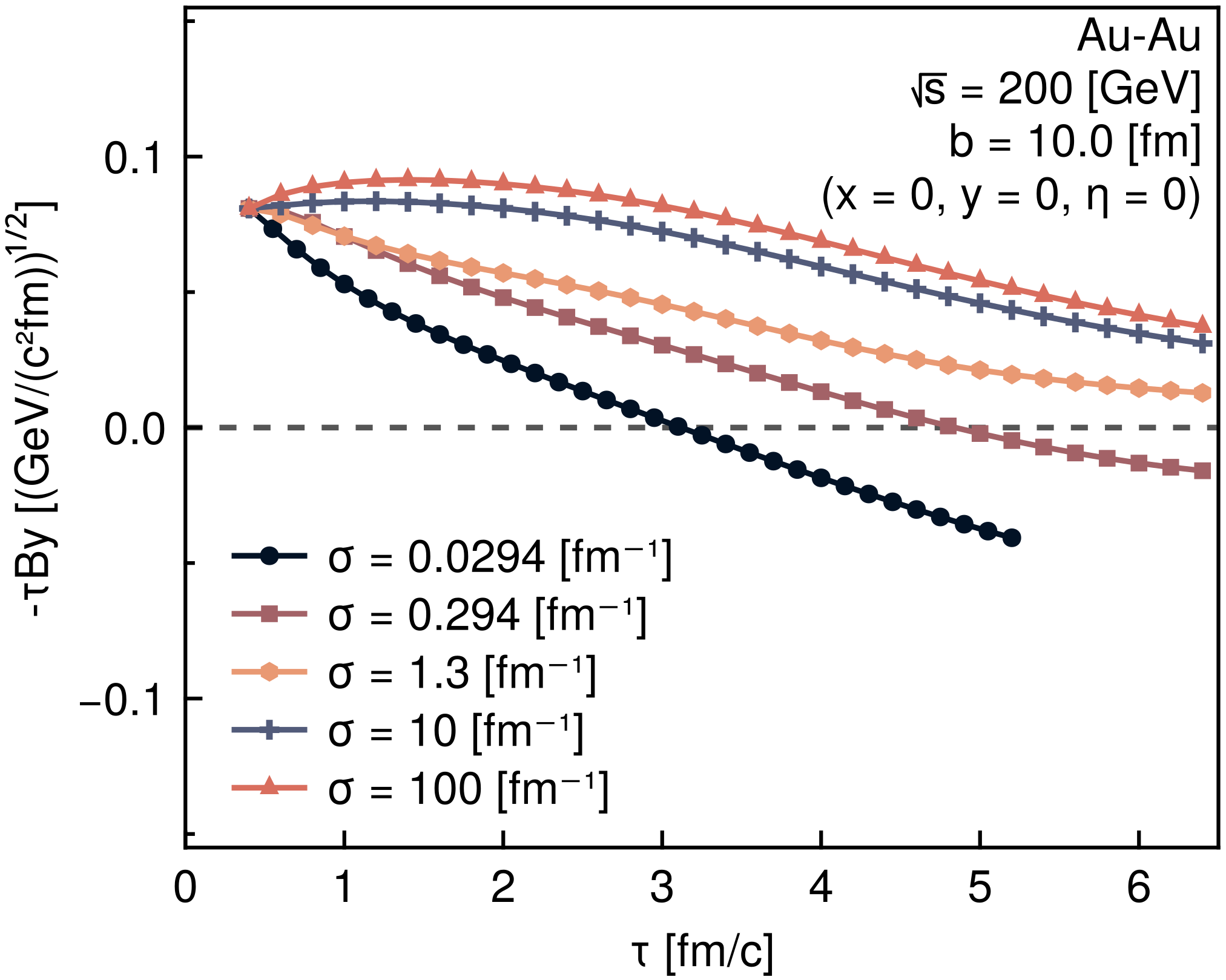}
    \caption{Time evolution of the $y$-component of the magnetic field at the center of the grid using our RRMHD model.
    We have assumed the electric conductivity of the plasma $\sigma$ is a constant in spacetime, and the spectators of the collision only act as an initial source for the field.
    The magnetic field's time evolution is similar to a non-linearly damped oscillator.
    For conductivity values between, $0 < \sigma < 1$ the magnetic field starts under damped and oscillates around zero.
    Larger values of $\sigma$ cause the magnetic field to be over damped, which means it never crosses zero.
    }
    \label{fig:comparedTimeEvo}
\end{figure}

Figure~\ref{fig:comparedTimeEvo} illustrates the impact of $\sigma$ on the time evolution of $B_y$ for our RRMHD model of QGP + EM fields.
Following the setup in Sec.~\ref{sec:numericalsetup} we prepare a peripheral Au-Au collision, with an impact parameter of $b=10$ fm, at the 
collision energy of $\sqrt{s_\text{NN}}=200$ GeV.
The RRMHD simulation was run from $\tau_0= 0.4$ fm/$c$ until the energy density for all grid cells below the freezeout value $\epsilon_f = 0.15 \text{ GeV/fm}^3$.
Focusing on the center of the grid $(x=0 \text{ fm}, y=0 \text{ fm}, \eta=0 )$, in Fig.~\ref{fig:comparedTimeEvo} we find a different time evolution for $\tau B_y$ that depends on the electric conductivity.
Larger values of the electric conductivity result in a longer lifetime of the magnetic field.
%The difference is independent of our choice for the initial conditions of the collisions and matches our expectations based on the simplified equations Eq.\eqref{eq:nonlinearBx}, Eq.\eqref{eq:nonlinearBy}, and Eq.\eqref{eq:nonlinearBz}.
Because we have assumed the initial conditions are the same, the point at the earliest time in Fig.~\ref{fig:comparedTimeEvo} is equal.
Additionally, from equations Eq.\eqref{eq:nonlinearBx}, Eq.\eqref{eq:nonlinearBy}, and Eq.\eqref{eq:nonlinearBz} we expect the magnetic field to behave as a non-linear oscillator.
In particular, values of $\sigma < 1\text{ fm}^{-1}$ result in oscillations around zero following under-damped behavior.
Whereas, those of $\sigma > 1\text{ fm}^{-1}$ never cross zero with $\tau B_y$ acting over-damped.

The reduction in differences between the results for larger sigma values, like between $\sigma = 10\text{ fm}^{-1}$ and $\sigma = 100\text{ fm}^{-1}$, is partly due to numerical diffusion overwhelming the physical diffusion.
This is a long-standing issue for resistive (relativistic) MHD models in astrophysics, where the conductivities can become relatively large, see Ref.~\cite{mattia_resistive_2023} as an example of a detailed analysis.
Here for each simulation, we keep the typical length-scale $L$ of the problem fixed and the order of the fluid velocities $u$ remains largely unchanged, but the conductivity increases by an order of magnitude.
This means the magnetic Reynolds number $S \equiv \sigma L |u|$ also increases by an order of magnitude for each conductivity.
When studying QGP, we limit the conductivity to be $\sigma < 5\text{ fm}^{-1}$.
This limit is justified by Lattice, pQCD, and AdS/CFT calculations and should keep the numerical diffusion smaller than the physical~\cite{hattori_electrical_2016,li_conductivities_2018,astrakhantsev_lattice_2020,aarts_electrical_2021,ghosh_electrical_2024,almirante_electrical_2024}.

\begin{figure}[bh]
    \includegraphics[width=\linewidth]{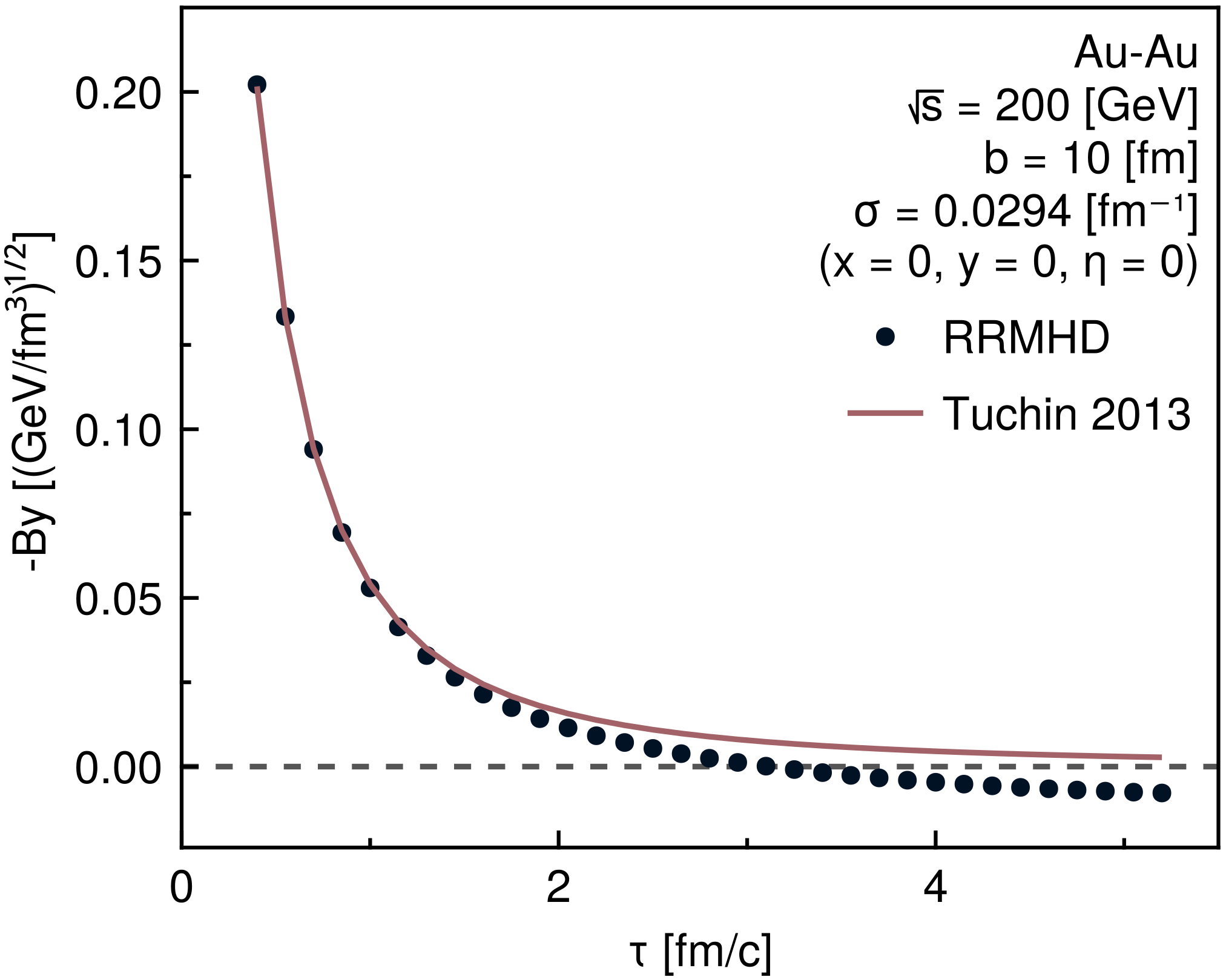}
    \caption{
        Time evolution of the magnetic field $B_y$ at the center of the grid $(x=0, y=0, \eta = 0)$ comparing our 
        %relativistic resistive magneto-hydrodynamic (RRMHD) model 
        RRMHD model to an analytic calculation of the field generated by the collision spectators.
        The first point is equal because the initial condition of the RRMHD model is the analytic solution.
        We attribute the differences between the models at later times to be due to the EM sources. The analytic solution only includes the collision spectators as a source, whereas the RRMHD model only includes QGP charges as a source.
    }
    \label{fig:analyticVsRRMHD}
\end{figure}

A useful feature of the analytic solutions used for the initial EM fields, see Appendix ~\ref{append:initialconditions} and Sec.~\ref{sec:numericalsetup}, is that it can be applied to all times.
As a comparison in Fig.~\ref{fig:analyticVsRRMHD} the analytic solution for $-B_y$ follows a similar time evolution as our RRMHD solution.
The deviations after $\tau\sim 1.25$ fm/$c$ between the analytic and RRMHD solutions can be attributed to different EM sources used in each calculation.
% At early times, both our RRMHD model and the analytic calculation only use the spectators as sources of the EM fields.
During the time evolution, our RRMHD model only considers QGP as a possible EM source $q$ in Eq.\eqref{eq:ohmslaw}.
Whereas, the analytic calculation includes all charges from the colliding ions, see Eq.\eqref{eq:initialEfield} and the discussion after it.
The difference between the two results may indicate our RRMHD model underestimates the EM fields, although the analytic solution is known to be an overestimation~\cite{tuchin_initial_2016,yan_dynamical_2023}.
At later times, we expect the charges of the spectators to have a limited effect because they have moved away from the collision region.

Overall, we can find general agreement about the nature of the magnetic field between our RRMHD model solutions and expectations.
The magnetic field rapidly decreases from its initial peak within a time frame of $\sim 2$ fm/$c$.
How the magnetic field decreases depends on the support of the plasma, with larger conductivity values increasing the life-time of the magnetic field, as seen in Fig.~\ref{fig:comparedTimeEvo}.
We expect this dependence to be reflected in the charge dependent directed flow $\Delta v_1$, because of changes to the plasma's electric charge density brought on by the Lorentz plus Faraday currents.

%%%% Results/Discussions part 2 %%%%
\section{RRMHD Results: Charge dependent directed flow}\label{sec:chargeDependentFlow}

Suggested as an observable for the EM fields, charge dependent directed flow $\Delta v_1$ can be a measure of the electric currents inside QGP~\cite{gursoy_charge-dependent_2018,gursoy_magnetohydrodynamics_2014}.
It is defined as the difference between the observed directed flows of positive electrically charged particles and negative electrically charged particles,
\begin{equation}\label{eq:differenceflow}
    \Delta v_1(Y) \equiv v^+_1(Y) - v^-_1(Y).
\end{equation}
Each charge dependent directed flow is calculated with the method~\cite{shen_iebe-vishnu_2016},
\begin{equation}\label{eq:chargeddirectedflow}
  v^\pm_1(Y) = \frac{\int p_T d p_T d \phi \cos \phi \frac{d N^\pm}{d Y p_T d p_T d \phi}}{\int p_T d p_T d \phi \frac{d N^\pm}{d Y p_T d p_T d \phi}},
\end{equation}
which is dependent on $Y$ the momentum-space rapidity and the hadron distribution is calculated using the Cooper-Frye formula of Eq.\eqref{eq:cooperfrye}~\cite{Cooper_1974,huovinen_particlization_2012}.
Finally, we numerically integrate over the transverse momentum $p_T$ and angle $\phi$ in Eq.\eqref{eq:chargeddirectedflow}.

For our model, the source of non-vanishing $\Delta v_1$ is a nonzero electric chemical potential $\mu_Q$ on the freezeout hypersurface, which modifies the local thermal distribution $f(x,p)$ of Eq.\eqref{eq:localdistribution}.
A nonzero chemical potential appears because locally the electric charge $q$ of the plasma is nonzero.
From Eq.\eqref{eq:ohmslaw} Ohm's law $q=-J^\mu u_\mu$, changes to the comoving charge density depend on the Lorentz and Faraday currents of Eq.\eqref{eq:restframeEfield}, which are proportional to the conductivity $\sigma$.

\begin{figure}[bh]
    \includegraphics[width=\linewidth]{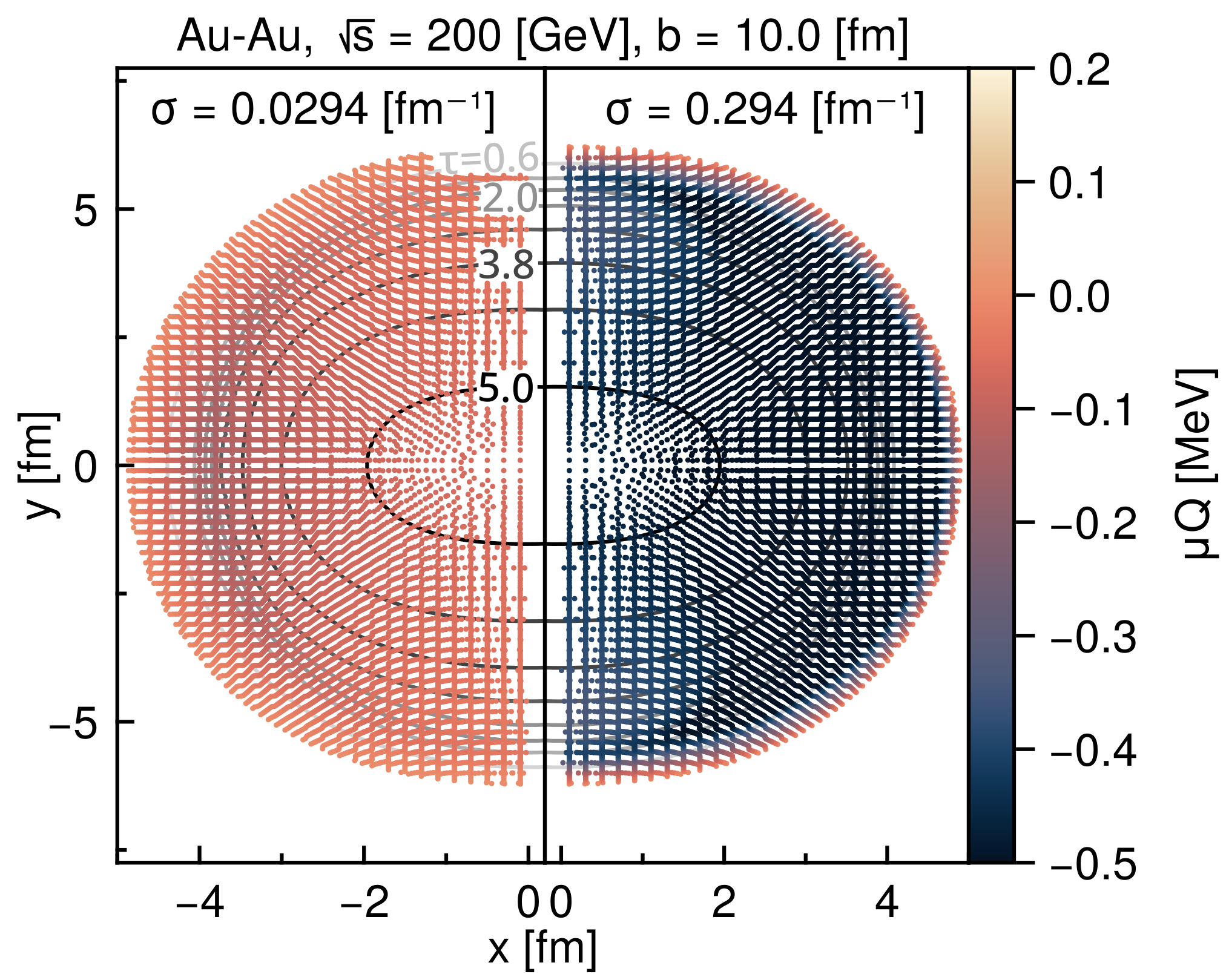}
    \caption{
        Electric chemical potential $\mu_Q$ that is accumulated on the freezeout hypersurface for $T_f = 0.140$ GeV and $ 0.2 > \eta > 0.0$.
        Contours are over constant proper times of $\tau > 0.4$ fm/$c$.
        The inner contours denote later times.
        Because we assumed the initial proton charges were equally distributed inside the ion, the transverse plane has reflection symmetry about zero on the $x$-axis and $y$-axis.
        The difference in magnitude of $\mu_Q$ between the panels is attributed to the value of QGPs electric conductivity~$\sigma$.
    }
    \label{fig:freezeoutmuQtransverse}
\end{figure}

Starting from an initial charge density of zero, Fig.~\ref{fig:freezeoutmuQtransverse} is a slice of the freezeout hypersurface at mid-rapidity in the transverse plane.
The contours are along constant proper times $\tau$ starting from the edge of the freezeout hypersurface.
The left panel is for a conductivity of $\sigma=0.0294\text{ fm}^{-1}$ and the right panel is for a larger conductivity of $\sigma=0.294\text{ fm}^{-1}$.
The differences in the panels illustrate that the magnitude of $\mu_Q$ depends on $\sigma$, larger conductivities result in larger magnitudes of $\mu_Q$.
This means the observable $\Delta v_1$ should depend on the conductivity of QGP via $\mu_Q$ and Eq.\eqref{eq:localdistribution} the local thermal distribution.

\begin{figure}[th]
    \includegraphics[width=\linewidth]{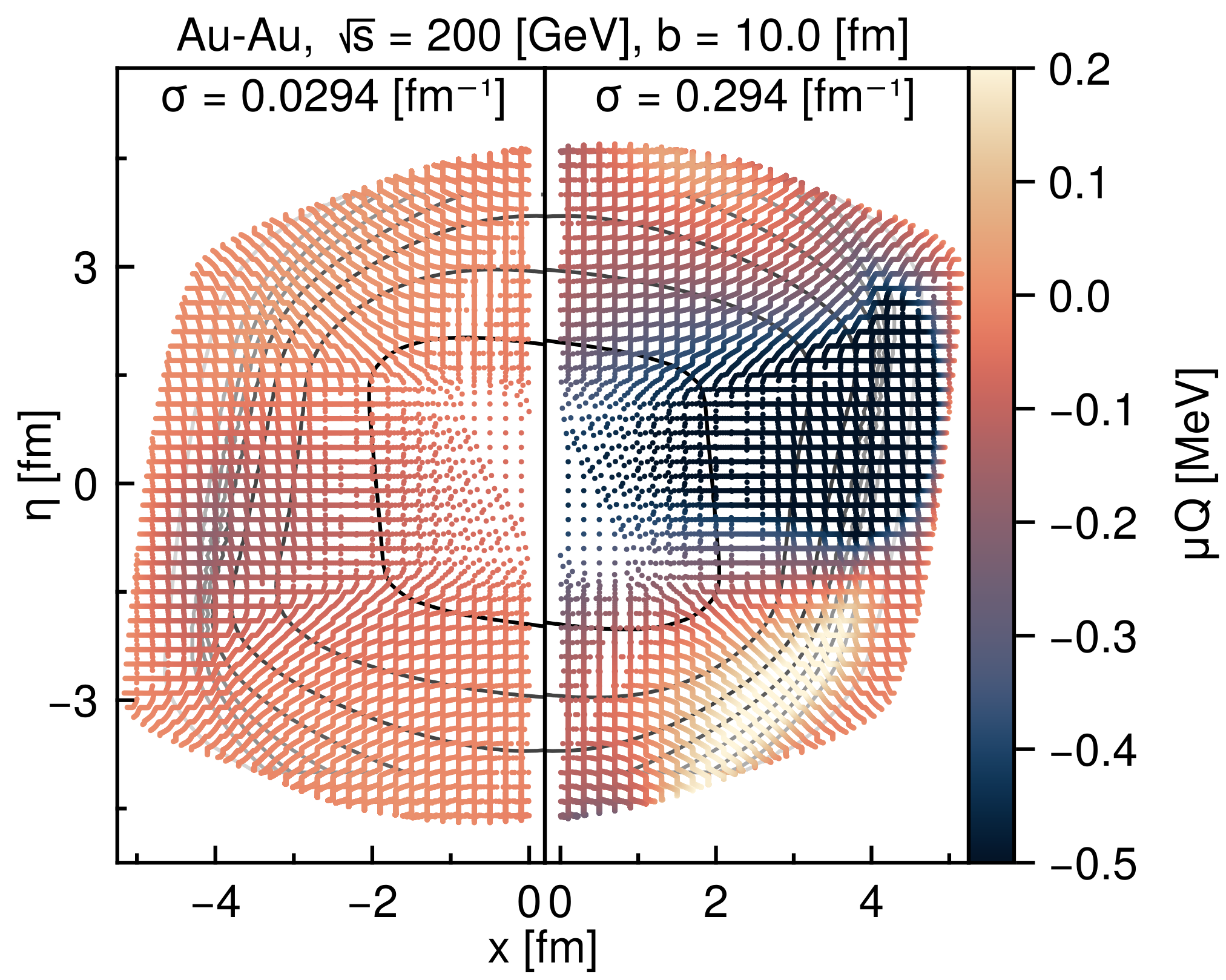}
    \caption{
        Longitudinal dependence of the electric chemical potential $\mu_Q$ that is accumulated on the freezeout hypersurface at $T_f = 0.140$ GeV and $0.2 > y > 0.0$.
        The contours are equivalent to the ones in Fig.\ref{fig:freezeoutmuQtransverse}, they are over constant proper times of $\tau > 0.4$ fm/$c$.
        For peripheral collisions, the magnetic fields in the longitudinal plane $\eta$ are asymmetric about zero~\cite{nakamura_relativistic_2023}.
        That leads to an asymmetry in $\mu_Q$ across $\eta$, which appears on the freezeout hypersurface.
        A 180 degrees rotation symmetry about the center $(0,0)$ exists because we assumed the initial proton changes were equally distributed inside the ions.
    }
    \label{fig:staticmuQ}
\end{figure}

In Fig.~\ref{fig:freezeoutmuQtransverse}, $\mu_Q$ has a reflection symmetry about $x=0$ and $y=0$ for that slice of the freezeout hypersurface.
In contrast, Fig.~\ref{fig:staticmuQ} illustrates an asymmetry in $\mu_Q$ across the longitudinal axis $\eta=0$ on the freezeout hypersurface.
This asymmetry is created in peripheral collisions, i.e., $b\neq 0$ fm, because the initial EM fields of the spectators break the reflection symmetry across $\eta = 0$~\cite{nakamura_directed_2023}.
Consequently, the charged directed flow of Eq.\eqref{eq:chargeddirectedflow} is different for positively and negatively charged final state hadrons.

\begin{figure*}[htb]
    \includegraphics[width=\linewidth]{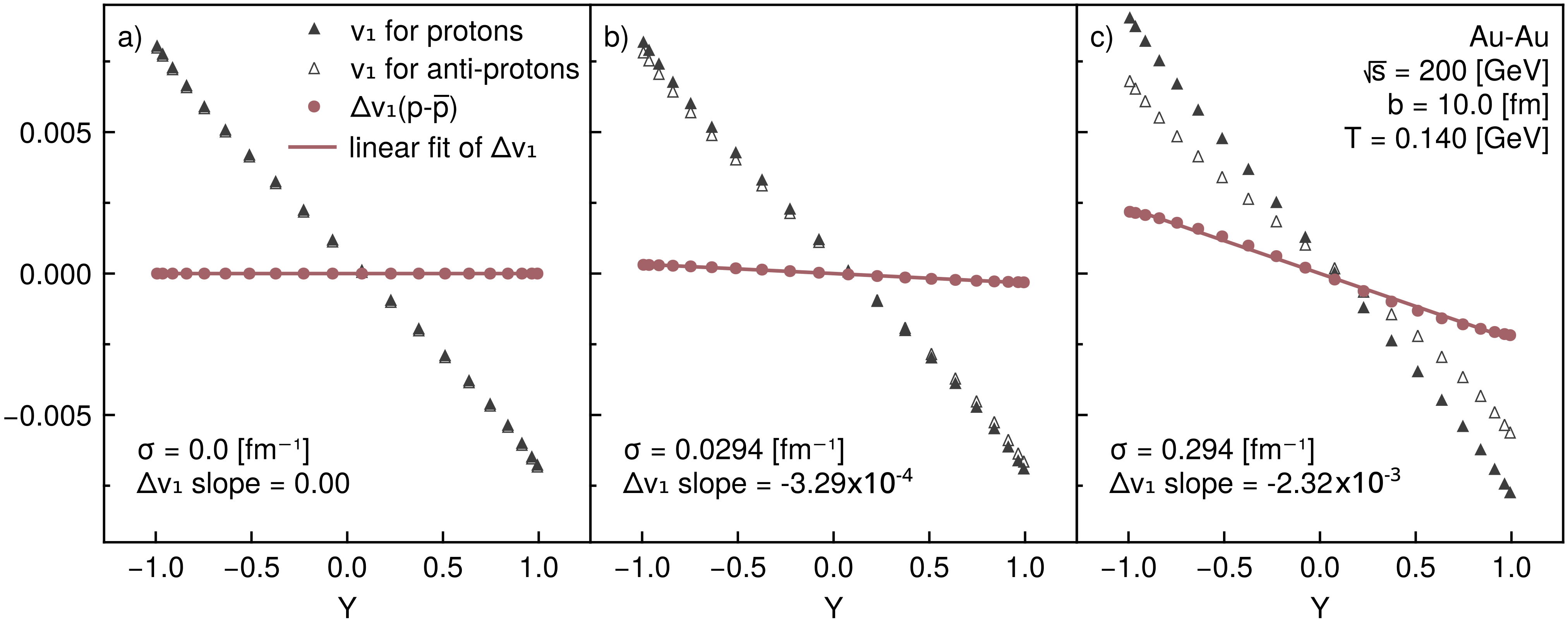}
    \caption{
        Charge dependent directed flow, calculated using our relativistic resistive magneto-hydrodynamic (RRMHD) model~\cite{nakamura_relativistic_2023}.
        In panel a, the conductivity of the plasma is zero, which results in protons and anti-protons having the same rapidity ($Y$) dependent directed flow $v_1$.
        In contrast, for panels b and c a separation of the proton, antiproton directed flow occurs for positive electric conductivities $\sigma > 0$.
        This separation appears because of a finite electric chemical potential on the freezeout hypersurface caused by the coupling between the electromagnetic fields and the plasma.
        The linear slope of the difference $\Delta v_1(p-\overline{p})$ is marked by a solid line.
        Negative values of the slope indicate the Faraday currents are greater than the Lorentz currents inside the plasma.
        We also vary the initial tilt of the energy density in the longitudinal direction and find it contributes to $\Delta v_1$ at a sub-leading level $\mathcal{O}(0.0001)$ to the slope.
    }
    \label{fig:protonDeltaV1}
\end{figure*}

An observable difference between the flow of protons and anti-protons is illustrated in Fig.~\ref{fig:protonDeltaV1}, where the conductivity of QGP is varied across the panels.
The rapidity~$Y$ dependence of (anti-)proton charge dependent directed flow Eq.\eqref{eq:chargeddirectedflow} separates in panels b and c.
That separation is captured by a nonzero difference $\Delta v_1 \neq 0$ from Eq.\eqref{eq:differenceflow}.
This is a similar result as the observations at the STAR experiment~\cite{star_observation_2024}.

In panel a of Fig.~\ref{fig:protonDeltaV1} for zero conductivity, $\sigma = 0 \text{ fm}^{-1}$, the difference between flows is zero because we have assumed the initial charge density is in equilibrium, i.e., $q_0(x,y,\eta)=0$.
This means the solution for Ohm's law in Eq.\eqref{eq:ohmslaw} is always zero.
By increasing the conductivity in panels b and c the proton and anti-proton directed flow separate.
This is expected from Ohm's law Eq.\eqref{eq:ohmslaw} and our discussions about Figs.~\ref{fig:freezeoutmuQtransverse} and \ref{fig:staticmuQ} that larger conductivities increase $\mu_Q$ on the freezeout hypersurface, which results in more significant differences between the charged directed flow.

The negative linear slope of $\Delta v_1$ in panels b and c of Fig.~\ref{fig:protonDeltaV1} is a sign that the Faraday current has a greater net effect than the Lorentz current of Eq.\eqref{eq:restframeEfield}.
That result is in agreement with Ref.~\cite{gursoy_charge-dependent_2018}.
This shows the approximation of the Faraday current as $\gamma B_y$ and the Lorentz current as $-\gamma u_z B_y$ at the beginning of Sec.~\ref{sec:magneticfield} holds for our RRMHD model.
Moreover, an order increase in the electric conductivity $\sigma$ brought a similar increase in the slope, $\Delta(dv_1/dy)$ suggesting this observable could be used to understand $\sigma$,  the electric conductivity of QGP.
Although, a more refined model is required to draw any quantitative conclusions.

%%%% Results/Discussions part 2a %%%%
\subsection{Centrality dependence of charge dependent flow}

In our previous work~\cite{nakamura_charge-dependent_2023}, $\Delta(v_1)$ of pions were studied instead of protons.
The result of that work proposed the electric conductivity of QGP should be $\sim \mathcal{O}(10^{-3})$ or smaller when fit to experimental data.
However, experimentally, the values of $\Delta(v_1)$ depend on the type of hadron (pions vs. protons) and the centrality of the collision~\cite{star_observation_2024,alice_probing_2020}.
The centrality dependence occurs because the EM field strength is proportional to the number of collision spectators, which means ultra-central collisions should have a smaller strength than peripheral collisions.
Then the decrease in the EM strength translates to smaller slopes of $\Delta v_1$.
This trend is reflected in the STAR data of $\Delta (dv_1 / dy)$ for protons and anti-protons between $20\%$ and $65\%$ centrality~\cite{star_observation_2024}.

\begin{figure}[htb]
    \includegraphics[width=\linewidth]{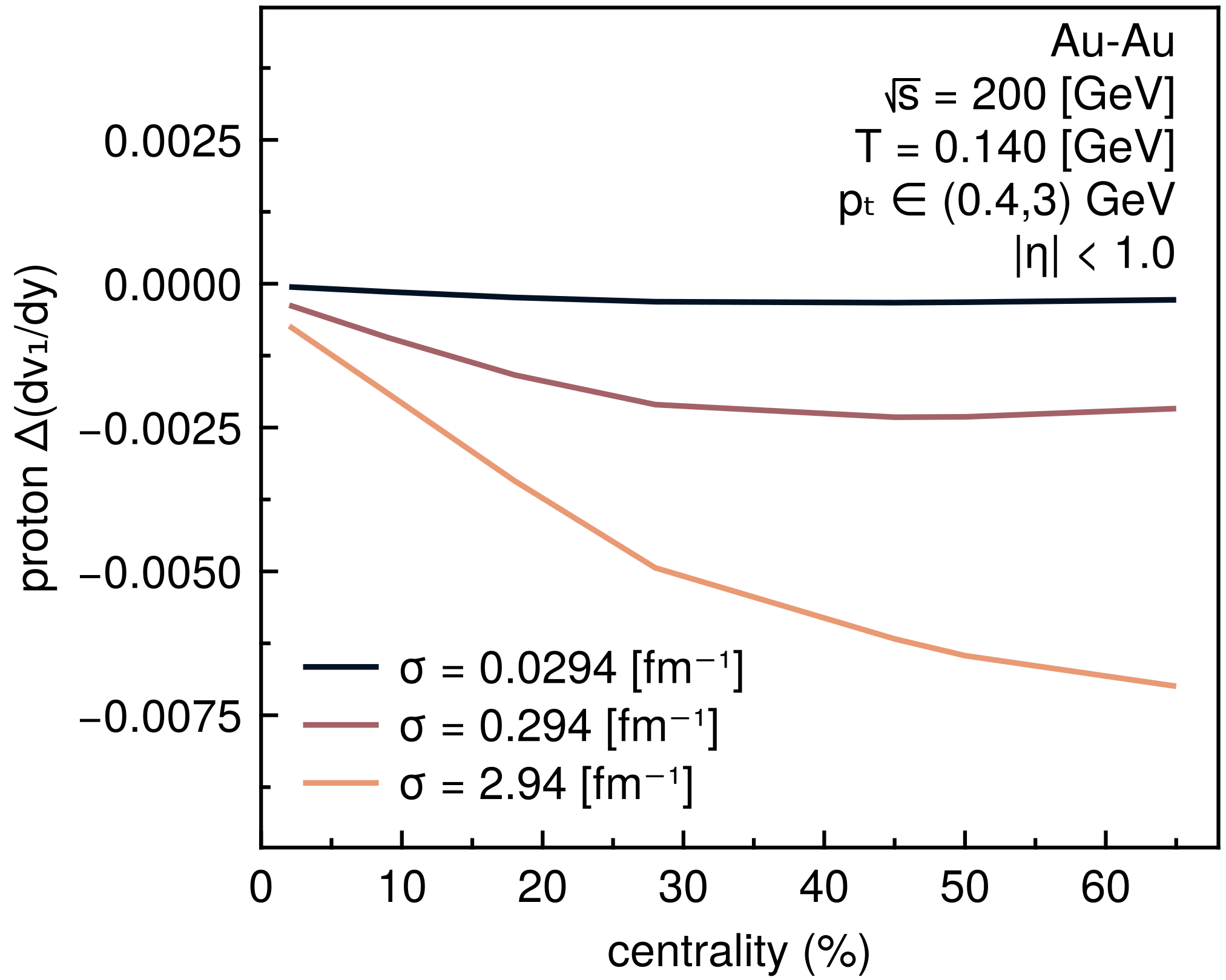}
    \caption{
        The slope of the difference in charge dependent directed flow for protons-antiprotons across collision centralities.
        From physical arguments, in central symmetric collisions the magnetic field strength is weaker, leading to smaller charged currents in the plasma.
        This is illustrated in our relativistic resistive magneto-hydrodynamic (RRHMD) model's predictions of a small slopes $\Delta(dv_1/dy)$ near 2\% centrality and larger slopes as centrality increases.
        At more peripheral collisions, our model requires a large electric conductivity $\sigma$ to predict similar slope values as STAR~\cite{star_observation_2024}. 
    }
    \label{fig:protonCentrality}
\end{figure}

Figure \ref{fig:protonCentrality} illustrates that our RRMHD model can capture how smaller centralities translate into a smaller $\Delta (dv_1/dy)$.
The dependence on electrical conductivity changes non-trivially across centralities, with a conductivity similar to lattice $\sigma = 0.0294 \text{ fm}^{-1}$ having the least change over centralities \cite{aarts_electrical_2021}.
In addition, the slope is always negative, which indicates the Faraday current always has a greater net effect than the Lorentz current in our model.
However, as was discussed in Ref.~\cite{gursoy_charge-dependent_2018}, the initial protons in the collision mean the plasma would have a net positive charge.
Until this point, we have considered the initial charge density to be zero, $q_0 = 0$.
The issue is initial dynamics of QGP have not been agreed upon, so a quantitative charge density distribution is difficult to establish.
If we consider a phenomenological model like Ref.~\cite{bozek_splitting_2022} for the baryon charge density and multiply it by the initial nuclei $n_Q/n_B$ to calculate the electric charge density, we can get a rough idea of the initial electric charge density $q_0$.
See Appendix~\ref{append:chargedensity} for a detailed discussion on the implementation.

For central collisions of $b = 2$ fm, by choosing $n_Q/n_B = 0.4$ based on $^{197}$Au, we see an increase in $\Delta (dv_1/dy)$,
\begin{equation}
    \Delta\left(\frac{dv_1}{dy}\right)\rvert_{\sigma = 0.0294} = -5.5307\times 10^{-5} \rightarrow 2.3513\times 10 ^{-4}.
\end{equation}
This demonstrates that an initial positive charge can result in positive slopes for $\Delta v_1$ with our RRMHD model.
The possibility of a positive slope also depends on the conductivity of QGP.
For example, the slope does not become positive for the largest conductivity value in Fig.~\ref{fig:protonCentrality} of $\sigma = 2.94\text{ fm}^{-1}$, 
\begin{equation}
    \Delta\left(\frac{dv_1}{dy}\right)\rvert_{\sigma = 2.94} = -7.3428\times 10^{-4} \rightarrow -3.7652\times 10^{-4}.
\end{equation}
Even though we started with an initial electric charge density $q_0 \neq 0$.

%%%% Conclusion %%%%
\section{Conclusions}\label{sec:conclusions}

We have demonstrated how our relativistic resistive magneto-hydrodynamic (RRMHD) model can be used to study the charge dependent directed flow $\Delta (dv_1/dy)$ across collision centralities.
The negative slope of $\Delta (dv_1/dy)$ was found to strongly depend on QGP's electric conductivity, $\sigma$, which is a unique feature of our model.
The slope of $\Delta (dv_1/dy)$ is caused by an electric charge imbalance on the freezeout hypersurface as illustrated in Figs.~\ref{fig:freezeoutmuQtransverse} and~\ref{fig:staticmuQ} in combination with the Cooper-Frye formula. 
%Eq.\eqref{eq:localdistribution}.
%As far as we are aware, 
This is the first work to establish such behavior and test is across centralities, which is 
shown in Fig.~\ref{fig:protonCentrality}.

In future work, we aim to improve our model in several ways.
Our results suggest that a complete modeling of $\Delta (dv_1/dy)$ requires a combination of baryon and electric charges.
This means incorporating the transported quark effects of Refs.~\cite{bozek_splitting_2022,guo_directed_2012,parida_baryon_2023,parida_charge_2025} with the electric conductivity effects of our RRMHD model.
It is possible to accomplish that incorporation using a phenomenological approach as described in Appendix~\ref{append:chargedensity}, or a more sophisticated model of the initial collision stages.
Moreover, a dedicated analysis of how different models of the initial stages change the initial EM fields + charge density would be interesting on its own.

Another important improvement is the inclusion of vortical effects on the magnetic field.
It has been established by experimental data with phenomenological models that collision participants and the resulting QGP have a large angular momentum.
Theoretical models suggest that angular momentum could positively modify the time evolution of the magnetic fields.
Incorporating that effect in our model should result in a better description of the electromagnetic fields.

The previous improvements focus on improving the accuracy of calculating the charge density $q$.
For a consistent calculation, we should use an equation of state (EoS) that includes the electric chemical potential during the time-evolution, for example~\cite{monnai_four-dimensional_2024}.
Considering this work found the charge density to be small on the freezeout hypersurface, see Figs.~\ref{fig:freezeoutmuQtransverse} and \ref{fig:staticmuQ}, the impact of changing the EoS during the time-evolution to include the electric chemical potential may also be small.

In this work, we have assumed the electric conductivity of QGP is a scalar.
But, dimensional analysis shows the conductivity should be temperature-dependent and dedicated calculations of the conductivity agree~\cite{aarts_electrical_2021}.
Additionally, for intense magnetic fields, the conductivity splits into a component parallel with the magnetic fields and a component perpendicular.
Considering both of those effects would change the conductivity from a scalar to a tensor.
A future investigation about how a tensor conductivity could affect experimental observables maybe interesting.

We have chosen a single freezeout temperature and then used the Cooper-Frye formula to calculate the final particle distributions.
This is quite an oversimplification of the final hadron dynamics that would be better described with an afterburner + EM field calculation.
Possible candidates are the kinetic solvers UrQMD~\cite{Bleicher:1999xi}, SMASH~\cite{SMASH:2016zqf}, or JAM~\cite{Nara:1999dz}, and ideally, we would like our model to be able to interface with all three.

Lastly, it has been well established that viscous hydrodynamics is required to explain the elliptical flow $v_2$ of final state hadrons.
Although, it may seem natural to include viscosity in our model, because of how it could modify the Lorentz current, we do not believe it to be as important as other improvements.
For example, our results agree with Ref.~\cite{gursoy_charge-dependent_2018} even though they use iEBE-VISHNU~\cite{shen_iebe-vishnu_2016} + a perturbative EM field calculation.

Even without those changes, we have demonstrated a relationship between the negative slope of $\Delta (dv_1/dy)$ and QGP's electric conductivity, $\sigma$ in Figs.~\ref{fig:protonDeltaV1} and~\ref{fig:protonCentrality}.
This contrasts with other works~\cite{bozek_splitting_2022,guo_directed_2012} that have focused on initial baryon charges and currents to explain the experimental data.
In addition, we also demonstrated how initial positive charge can also cause positive slopes for $\Delta v_1$, as was originally discussed in Ref.~\cite{gursoy_charge-dependent_2018}.
All of this work suggests an understanding of charge dependent directed flow at STAR~\cite{star_observation_2024} and ALICE~\cite{alice_probing_2020} calls for a consideration of QGP's electric conductivity.

\begin{acknowledgments}
We thank Azumi Sakai for useful discussions and a thorough review of the initial manuscript.
Furthermore, we thank Hidetoshi Taya and Akihiko Monnai for insightful discussions about early versions of this work.
The numerical computation in this work was carried out at the Yukawa Institute Computer Facility.
N.J.B. is grateful for the financial support from Hiroshima University's presidential discretionary fund and Start-Up program fund.
This work was also supported by JSPS KAKENHI Grant Numbers, JP20K11851, JP20H00156, JP24K07117 (T.M.), JP20H00156, JP20H11581 (C.N.), JP21H04488, JP24K00672, and JP24K00678 (H.R.T.) and by the World Premier International Research Center
Initiative (WPI) under MEXT, Japan (C.N.).
Perceptually uniform color maps are used in this study to prevent visual distortion of the data and to better serve those with color vision deficiency~\cite{crameri_misuse_2020}.
\end{acknowledgments}

%%%% Appendix %%%%
\appendix
\section{Initial electromagnetic fields}\label{append:initialconditions}
For the work, we calculated the initial EM fields using the results of Ref.~\cite{tuchin_time_2013} which we briefly summarize.
Starting from the EM fields produced by a single charge $e$ only moving along the beam axis $z$ with a fixed velocity $v$, the in-medium Maxwell equations are solved.
By taking the relativistic limit $\gamma=1/\sqrt{1-v^2} \gg 1$ and accounting for a conductive medium $\gamma \sigma b \gg 1$, the solutions to Maxwell equations become,
\begin{gather} \label{eq:initialEfield}
    E_r = B_\phi = \frac{e}{2\pi}\frac{b\sigma}{4(t\pm v/z)^2}\exp{\left(-\frac{b^2\sigma}{4(t\pm v/z)}\right)},
    \\
    E_z = - \frac{e}{4\pi}\frac{(t\pm v/z)-b^2\sigma}{\gamma^2 (t\pm v/z)^3}\exp{\left( -\frac{b^2\sigma}{4 (t\pm v/z)} \right)},
\end{gather}
in cylindrical coordinates.
Those are the same equations previously used in Ref.~\cite{nakamura_directed_2023}, but we have suppressed the $\hbar c$ factors and the dimensions of the EM fields should be understood implicitly to be $\sqrt{\text{GeV}/\text{fm}^3}$.

We account for the multiple charges of heavy-ions like ${}^{197}$Au in Tab.~\ref{tab:initialparameters} by diffusing the total charge equally inside the ion $\rho=Z/\left(\frac{4}{3}\pi R^3\right)$.
Here $Z$ is the number of protons inside the ion and $R$ is the ion's radius.
Then we integrate over the contribution of all the charge densities to arrive at the EM field configuration created by a single ion moving along the $z$ axis, which is step 2 of Sec.~\ref{sec:numericalsetup}.
The same method is applied to the other ion involved in the collision, but the sign of the velocity is reversed $-v$~\cite{tuchin_time_2013}.

Because the charge is equally distributed inside the ion, we refer to this method as smooth initial conditions.
This is contrary to fluctuating initial conditions, where the charge density can vary inside the ion.

\section{Effects from an initial charge density}\label{append:chargedensity}
We include a rough estimate of the initial charge density based on the phenomenological ansatz used to calculate and initial baryon number in Ref.~\cite{bozek_splitting_2022}.
This involves reusing the wounded nucleon's weight profile of the tilted optical Glauber model.
Because this is a rough estimate, we assume an initial charge density of,
\begin{equation}\label{eq:initialcharge}
    \rho_q(x_\perp, \eta;b)=Z_q\frac{T_-(x_\perp; b)f_-(\eta)+T_+(x_\perp; b)f_+(\eta)}{T_-(0)f_-(0)+T_+(0)f_+(0)}H(\eta).
\end{equation}
The terms $T_\pm(x_\perp; b)$ are the thickness functions of the individual nuclei shifted in the transverse plane by the impact parameter $b/2$,
\begin{equation}
    T_\pm(x_\perp; b) = A T(x\pm b/2,y) \left(1-\left(1-\sigma_\text{NN}T(x\mp b/2,y)\right)^A\right),
\end{equation}
where $T(x\pm b/2,y)$ are the normalized nuclei Fermi distributions and $A$ is the nuclear mass.
For this work, we have used a radius of $r_0 = 6.38 \text{ fm}$, a width of $\delta = 0.535 \text{ fm}$, and a nuclear saturation density of $\rho_0 = 0.17 \text{ fm}^{-3}$ for $^{197}$Au.
The functions $f_\pm(\eta)$ in Eq.\eqref{eq:initialcharge} impose a tilt in the longitudinal direction and $H(\eta)$ define a longitudinal profile~\cite{bozek_directed_2010}.
For the initial baryon number calculation in Ref.~\cite{bozek_splitting_2022}, an additional longitudinal weighting is preformed with $H_B(\eta)$ that has two free parameters $\eta_B$ and $\sigma_B$.
We ignore $H_B(\eta)$ in this work because fixing those parameters requires solving the hydrodynamic evolution with a nonzero baryon current such that the ratio of transported and produced baryons at central rapidities can be reproduced.
We do include a ratio parameter $Z_q$ between the initial number of baryons vs. number of protons, which acts as a conversion from the initial baryon charge to the initial electric charge.

\section{Dependence on the strength of the initial electromagnetic fields}\label{append:initialEMfields}
We mentioned in the second bullet point of Sec.~\ref{sec:magneticfield} the pre-equilibrium dynamics after the collision could change the configuration of the initial EM fields.
An example of a change is the results in Ref.~\cite{yan_dynamical_2023}, which uses a Kinetic model for the pre-equilibrium dynamics.
Their results suggest the EM fields initially decay with a time dependence of $\lvert eB\rvert\sim\tau^{-3}$, like in a vacuum~\cite{Hattori:2016emy}.
Then the decay smoothly transition to a time dependence of $\lvert eB\rvert\sim\tau^{-1}$, which is an in-medium like decay behavior~\cite{deng_event-by-event_2012,Roy:2015kma}.
In contrast, the initial condition we used assumes the EM fields always decay in a conducting medium~\cite{tuchin_time_2013}.
This indicates our initial conditions overestimate the strength of the EM fields.

\begin{figure}[ht]
    \includegraphics[width=\linewidth]{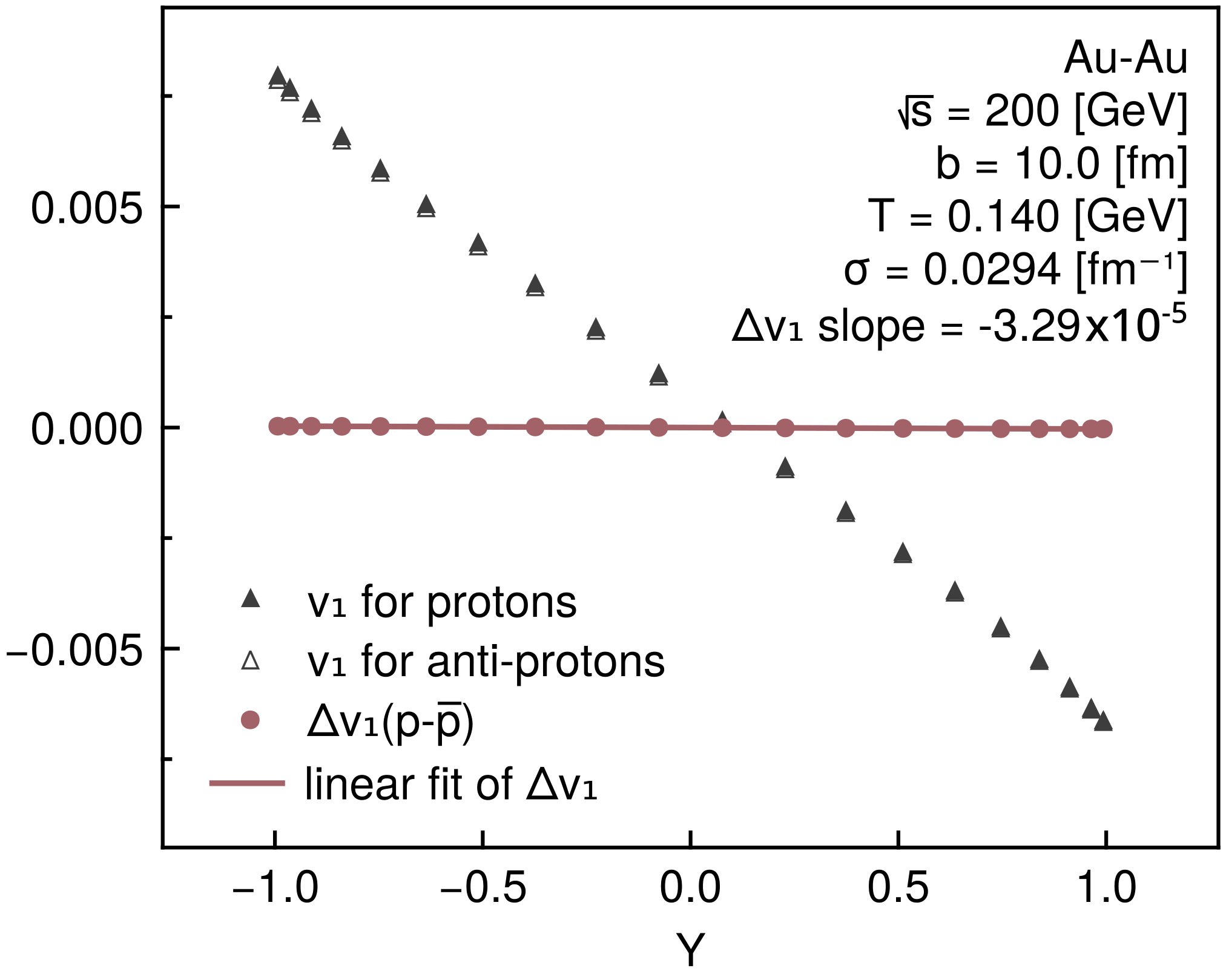}
    \caption{
        Charge dependent directed flow, calculated using our relativistic resistive magneto-hydrodynamic (RRMHD) model~\cite{nakamura_relativistic_2023}.
        Compared to panel b of Fig.\ref{fig:protonDeltaV1}, the initial EM fields have been reduced to $10\%$ of the original values.
        All other parameters are the same as panel b.
        The resulting $\Delta v_1$ slope also decreased on the order of $10\%$ relative to the slope in panel b.
    }
    \label{fig:reducedEMfields}
\end{figure}

To roughly understand how that overestimation could affect our results, we have scaled down the strength of the initial EM fields to $10\%$ of the original values~\cite{tuchin_time_2013}.
Then we reproduced panel b) of Fig.~\ref{fig:protonDeltaV1}, for which the conductivity was similar to lattice QCD values $\sigma = 0.0294\text{ fm}^{-1}$.
The result is illustrated in Fig.~\ref{fig:reducedEMfields}, the slope of $\Delta v_1$ becomes reduced by an order of magnitude from $\mathcal{O}(10^{-4})$ to $\mathcal{O}(10^{-5})$.
This is approximately proportional to the same reduction of the initial EM fields of $10\%$.
Although this suggests our results are equally sensitive to the electric conductivity of QGP $\sigma$ and the initial EM field intensities, more robust modeling of the pre-equilibrium dynamics is required to draw any firm conclusions.

\bibliography{conductivity.bib}% Produces the bibliography via BibTeX.

\end{document}